\documentclass[10pt,conference,letterpaper]{IEEEtran}
\IEEEoverridecommandlockouts
\usepackage{cite}
\usepackage{url}
\usepackage{stfloats}
\usepackage[font=footnotesize]{caption}
\usepackage{amsmath,amssymb,amsfonts}
\usepackage{bm}
\usepackage{algorithm}
\usepackage{algpseudocode}
\usepackage{graphicx}
\usepackage{textcomp}
\usepackage{xcolor}
\usepackage[utf8]{inputenc}
\usepackage[normalem]{ulem}
\usepackage{soul}
\usepackage{ulem}
\usepackage{hyperref}
\usepackage[capitalise]{cleveref}

\usepackage[letterpaper, left=0.625in, right=0.625in, bottom=1in, top=0.75in]{geometry}

\usepackage{acronym}
\acrodef{rsm}[RSM]{receive spatial modulation}
\acrodef{adc}[ADC]{analog-to-digital converters}
\acrodef{ris}[RIS]{reconfigurable intelligent surfaces}
\acrodef{zf}[ZF]{zero-forzing}
\acrodef{ras}[RAS]{receiver antenna selection}
\acrodef{ara}[ARA]{active receiver antenna}
\acrodef{aras}[ARAs]{active receiver antennas}
\acrodef{dl}[DL]{downlink}
\acrodef{snr}[SNR]{signal-to-noise ratio}

\DeclareUnicodeCharacter{2061}{}
\newcommand{\Hbu}{\mathbf{H}_{bu}}

\newcommand{\Hbr}{\mathbf{H}_{br}}

\newcommand{\Hru}{\mathbf{H}_{ru}}

\newcommand{\phiM}{\pmb{\varphi}}

\title{Beyond Diagonal RIS-assisted MIMO Transmission: Beamforming Gain and Capacity Optimization\\
\thanks{This work has been carried within the framework of the I+D+i project 6-SENSES (PID2022-138648OB-I00) funded by MICIU/AEI/10.13039/501100011033 and ERDF/EU and by FEDER-UE, ERDF-EU A way of making Europe, and the grants 22CO1/008248 and 2021 SGR 01033 (AGAUR, Generalitat de Catalunya).}}

\author{\IEEEauthorblockN{Ainna Yue Moreno-Locubiche, Josep Vidal}
\IEEEauthorblockA{\textit{Dept. of Signal Theory and Communications, Universitat Politècnica de Catalunya - BarcelonaTech (UPC), Spain} \\
\{ainna.yue.moreno, josep.vidal\}@upc.edu}}
\usepackage{subcaption}
\begin{document}
\maketitle
\begin{abstract}
Reconfigurable Intelligent Surfaces (RIS) have emerged as a transformative technology in wireless communications, offering unprecedented control over signal propagation. This study focuses on passive beyond diagonal reconfigurable intelligent surface (BD-RIS), which has been proposed to generalize conventional diagonal RIS, in Multiple-Input Multiple-Output (MIMO) downlink (DL) communication systems. We compare the performance of transmit beamforming (TxBF) and MIMO capacity transmission with waterfilling power allocation in the millimeter wave (mmWave) band, where propagation primarily occurs under line-of-sight (LOS) conditions. In the lack of closed-form expressions for the optimal RIS elements in either case, our approach adopts a gradient-based optimization approach requiring lower complexity than the solution in \cite{santamaria_c_bdris}. Numerical results reveal that BD-RIS significantly outperforms traditional diagonal RIS in terms of spectral efficiency and coverage.
\end{abstract}


\section{Introduction} \label{Intro}
The \ac{ris} have emerged as a transformative technology in wireless communications, promising enhanced performance with reduced power consumption and cost-effective architecture. An \ac{ris} consists of numerous reconfigurable reflective passive elements capable of manipulating the phase of electromagnetic waves to direct signals toward intended receivers. This capability positions RIS as a key enabler for future 6G networks \cite{Fang2023LowComplexity, Sun2023NewModel, Li2024Wideband}.

Conventional passive RIS architectures typically employ diagonal phase shift matrices (D-RIS), a constraint that limits the capacity for advanced signal manipulation and propagation environment shaping. Recently, the concept of Beyond Diagonal RIS (BD-RIS) has been introduced, leveraging inter-element connections to achieve higher degrees of freedom in electromagnetic wave control \cite{Nerini2023GraphTheory, Li2023ChannelEstimation}. BD-RIS has shown significant promise in improving coverage and energy efficiency in Multiple-Input Multiple-Output (MIMO) systems.

Recent studies have explored aspects of BD-RIS such as beamforming optimization \cite{Fang2023LowComplexity}, channel estimation \cite{Li2023ChannelEstimation}, and power allocation schemes \cite{Zhou2023OptimizingPower}. However, gaps remain in understanding the joint optimization of transmit precoding and BD-RIS coefficients, under single mode or multi-mode transmissions \cite{Santamaria2024SNRMaximization}. Addressing these gaps is critical to unlock the full potential of BD-RIS in enabling enhanced propagation channel control for next-generation communication systems. In this respect, our contribution is in providing answers to these questions:
\begin{itemize}
    \item How much spectral efficiency and coverage gains can BD-RIS provide over D-RIS architectures when integrated with beamforming or multimode MIMO transmissions in the presence of non-negligible direct links?
    \item How can gradient-based algorithms be effectively utilized to jointly optimize transmit beamforming and BD-RIS coefficients in single user MIMO downlink (DL) systems?
\end{itemize}

\subsection{Literature Review}
To cope with the inherent limitations of passive diagonal RIS, Beyond Diagonal RIS (BD-RIS) have been proposed, whereby an $N_r$-port reciprocal network where each port is
connected to all other by a reconfigurable reactance. This entails that the signal impinging on each RIS element can be divided, phase-shifted, and emitted from a multitude of other elements, thus enabling more complex and adaptive electromagnetic control \cite{Nerini2023GraphTheory, Li2024Wideband}.

The integration of BD-RIS in MIMO systems, particularly in DL scenarios, has garnered significant attention. Research in \cite{Li2024Wideband} provided a circuit-based model for BD-RIS in wideband OFDM systems, illustrating substantial improvements in signal-to-noise ratio (SNR) and interference mitigation. Similarly, \cite{Santamaria2024SNRMaximization} highlighted the role of BD-RIS in maximizing SNR in both single and multi-antenna configurations. In \cite{10155675}, closed-form solutions are proposed for the single user SISO case. For the single-mode MIMO case, a closed form solution is derived when the direct link is disregarded. Otherwise, they propose a suboptimal alternate maximization. The multi-mode transmission is not covered.

Transmit beamforming has been extensively studied in the context of RIS. The study by \cite{Fang2023LowComplexity} introduced a low-complexity beamforming design tailored for BD-RIS-assisted multi-user networks, demonstrating improved energy efficiency compared to diagonal RIS systems. Furthermore, \cite{Zhou2023OptimizingPower} explored joint optimization of beamforming and RIS configurations to optimize energy efficiency and power consumption.

Despite these advancements, there remains a need for an optimization framework capable of multi-mode MIMO transmission. While waterfilling algorithms for power allocation have long been a cornerstone of resource optimization in wireless communications, integrating them with BD-RIS is a challenge due to the non-convexity of the problem. \cite{santamaria_c_bdris} derives a solution based on a mixed alternate optimization and gradient computation on a lower bound of the BD-RIS MIMO capacity problem. Recently, \cite{10999443} has provided a closed form solution, but based on simplified assumptions: removing both the direct link and the symmetry constraint on the RIS scatter matrix.

To solve the general SU-MIMO case both for transmit beamforming and multi-mode transmission, we adopt a fresh gradient-based methodology founded on complex matrix calculus. This was used in \cite{10437329} in the context of D-RIS MIMO single-user receive spatial modulation (RSM) transmission for rank-deficient channels, where spectral efficiency was boosted through the RIS ability to improve channel matrix rank.

\subsection{Organization} The paper is structured as follows. Section \ref{Channel model including RIS} describes the single user system model and problem formulation. Sections \ref{Optimizing BD-RIS in Beamforming Transmission} and \ref{Optimizing BD-RIS in Multimode MIMO Transmission} describe the optimization problems for the transmit beamforming and the MIMO capacity optimization respectively. Section \ref{Gradient ascend iteration} describes the gradient-based solver proposed for the BD-RIS coefficients. Section \ref{Results} contains the results.

\subsection{Notation}
Boldface lower- and upper-case characters denote vectors and matrices respectively: $\mathbf{x}\in \mathbb{C}^{N\times 1}$ and $\mathbf{X}\in \mathbb{C}^{K\times N}$ are a vector of size $N$ and a matrix of size $K\times N$ respectively, with complex entries. The superscripts $(\cdot)^T$ and $(\cdot)^H$ represent the transpose and conjugate transpose. $\mathrm{Tr}\{\mathbf{X}\}$ is the trace of $\mathbf{X}$. $\left[\mathbf{X}\right]_{i,j}$ refers to the $(i,j)$th element of $\mathbf{X}$. $\text{vec}(\mathbf{X})$ rearranges the elements of  matrix $\mathbf{X}$ column-wise. The operator $\text{diag}(\mathbf{x})$ is the $N\times N$ diagonal matrix whose entries are the $N$ elements of vector $\mathbf{x}$. $\operatorname{Im}(\mathbf{x})$ stands for the imaginary part of vector $\mathbf{x}$.

\section{Channel Model Including RIS} \label{Channel model including RIS}
\begin{figure}
\begin{center}
\includegraphics[scale=0.2]
{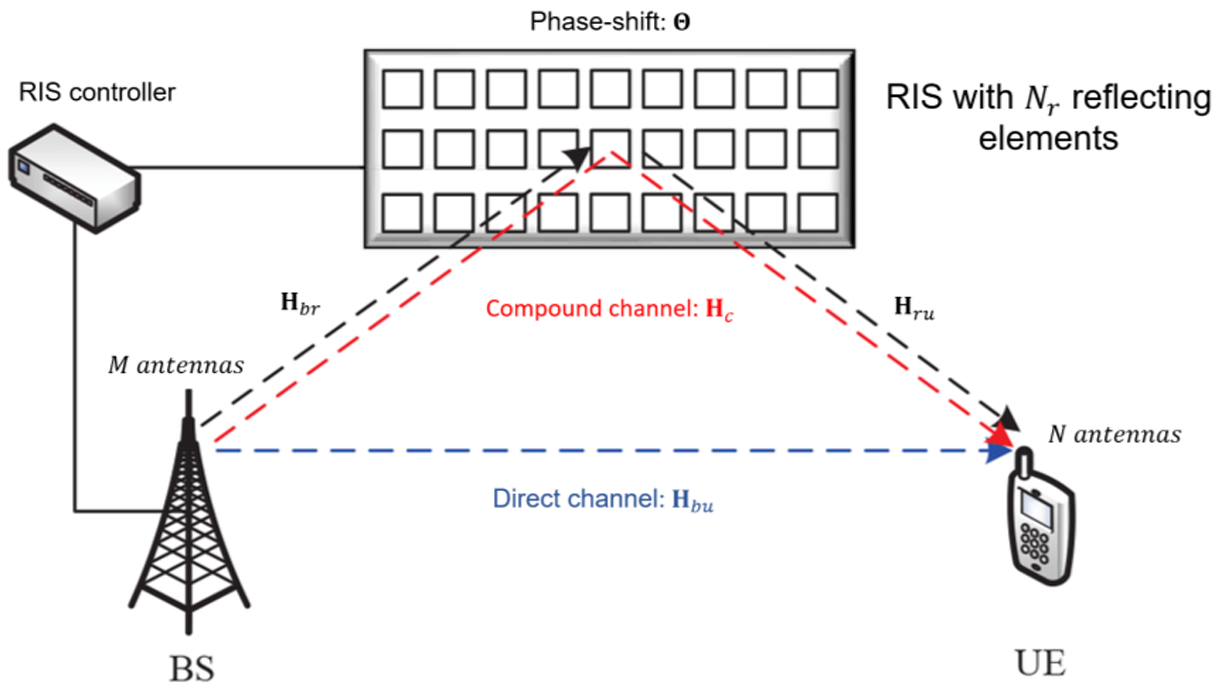}
    \caption{Channel model for the \ac{ris}-assisted DL transmission.}
    \label{fig:channel_model}
\end{center}
\end{figure}

Consider the single user MIMO scenario with the presence of a RIS whose all \(N_r\) elements are purely passive reflectors. The MIMO channel \(\mathbf{H}\) as shown in Fig. \ref{fig:channel_model} is expressed as:
\begin{equation}
   \label{eq:channel_model} 
   \mathbf{H}=\Hbu+\mathbf{H}_c \in \mathbb{C}^{K\times M}.   
\end{equation}
\(\Hbu  \in \mathbb{C}^{K\times M}\) contains the direct channel between the $M$-antenna BS and the $K$-antenna user. \(\mathbf{H}_c\) is the compound channel between the BS, RIS, and the user equipment (UE):
\begin{equation}
    \label{eq:compund_channel}
    \mathbf{H}_c=\Hru\mathbf{\Theta}\Hbr,
\end{equation}
where \(\Hbr\in\mathbb{C}^{N_r\times M}\) contains the channel gains between the BS and the \ac{ris}, \(\Hru\in\mathbb{C}^{K\times N_r}\) contains the channel gains between the \ac{ris} and the user, and $\boldsymbol{\Theta}$ is the scatter matrix that contains the reflection coefficients at the RIS.
The channel coefficients for the BS-UE link are written as:
\begin{equation}
    \label{eq:direct_channel_su}      [\mathbf{H}_{bu}]_{k,l}=\sqrt{\beta_{bu,k,l}}\exp{\left(-j \frac{2\pi f_c}{c}d_{k,l}\right)}
\end{equation}
where $f_c$ is the carrier frequency, $d_{k,l}$ is the distance between the $l$-th antenna at the BS and the $k$-th antenna at the UE, and $c$ is the speed of light. Likewise, the elements of $\mathbf{H}_{ru}$ and \(\Hbr\) are given by:
\begin{equation}
    \label{eq:channel_ru}
        [\mathbf{H}_{ru}]_{k,i}=\sqrt{g_{ru,k,i}}\exp{\left(-j \frac{2\pi f_c}{c}d_{k,i}\right),}
\end{equation}
\begin{equation}
    \label{eq:channel_br}
        [\Hbr]_{i,l}=\sqrt{g_{br,i,l}}\exp{\left(-j \frac{2\pi f_c}{c}d_{i,l}\right)},
\end{equation}
where $d_{k,i}$ is the distance between the $i$-th element of the RIS and the $k$-th antenna at the user, $d_{i,l}$ is the distance between the $i$-th element at the RIS and the $l$-th BS antenna, and {$\beta_{bu,k,l}$}, $g_{ru,k,i}$ and $g_{br,i,l}$ are pathloss components that will be discussed in Sec. \ref{Results}. We assume perfect channel state knowledge all through the paper.

In conventional passive D-RIS models, diagonal phase shifts are assumed: $\boldsymbol{\Theta} = \text{diag}(e^{j\phi_1}, e^{j\phi_2}, \dots, e^{j\phi_{N_r}})$. In BD-RIS, coupling between elements is introduced, enabling further degrees of freedom in the design of waveform shaping for beam focusing, multipath control, or spatial diversity. The scatter matrix $\boldsymbol{\Theta}$ is assumed to be symmetric due to electromagnetic reciprocity. Physically, this means the RIS imparts identical phase shifts to signals in both impinging directions, which is a fundamental property of lossless reciprocal scatterers. On the other hand, if the RIS is a passive device, a unitary matrix ensures that it does not alter the total reflected power \cite{Shen2022Modeling}. The feasibility set for $\mathbf{\Theta}$ is therefore:
\begin{equation}
\mathcal{T}=\{\mathbf{\Theta}|\mathbf{\Theta}^H\mathbf{\Theta}=\mathbf{I}_{N_r},\mathbf{\Theta}=\mathbf{\Theta}^T\}
\end{equation}

In the following sections we study the gradient-based joint optimization of transmit precoders and RIS coefficients for beamforming and multi-mode MIMO transmissions.

\begin{figure}
\begin{center}
\includegraphics[scale=0.2]
{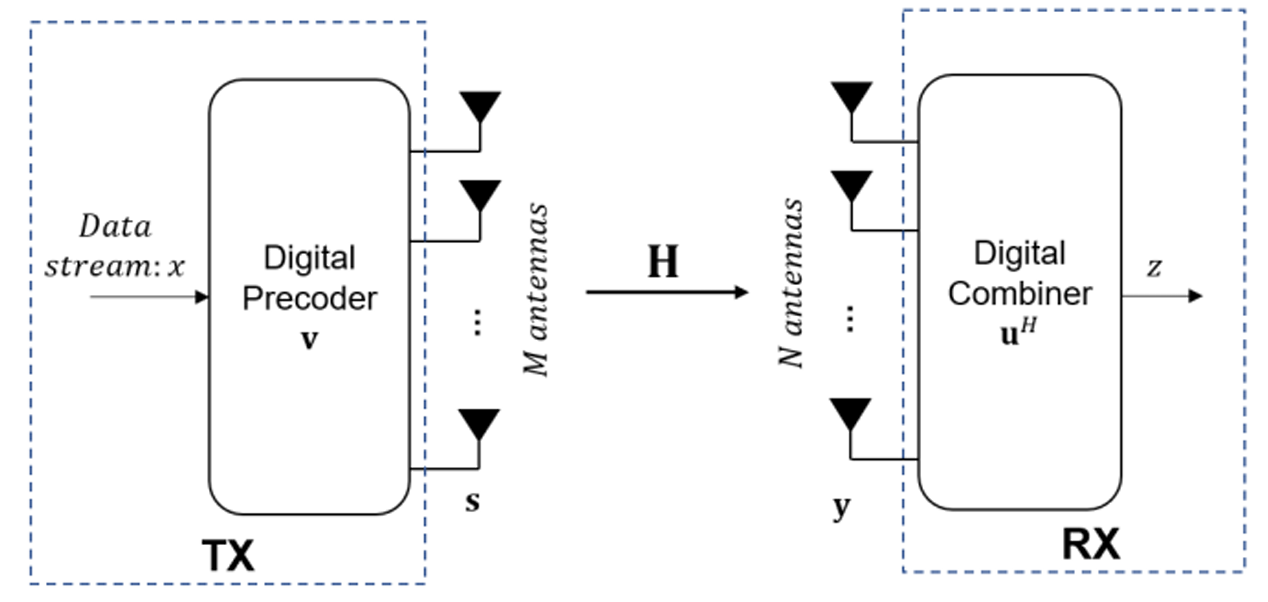}
    \caption{Block diagram of a MIMO system with transmit beamforming.}
    \label{fig:TxBF_model}
\end{center}
\end{figure}

\section{BD-RIS in Transmit Beamforming}
\label{Optimizing BD-RIS in Beamforming Transmission}
Consider Fig. \ref{fig:TxBF_model} where $\mathbf{H}$ in    \eqref{eq:channel_model} is the downlink channel, with $M$ antennas at the transmitter and $K$ antennas at the receiver and information is conveyed in a single channel mode. The received signal vector is given by:
\begin{equation}
    \mathbf{y} = \mathbf{H} \mathbf{v}s + \mathbf{n} \in \mathbb{C}^{K \times 1}
\end{equation}
where $s$ is the unit-power transmitted data stream, $\mathbf{v}$ is the beamforming vector and $\mathbf{n}$ is the additive receiver Gaussian noise of zero mean and covariance $\sigma^2\mathbf{I}$. The optimization problem is formulated as:
\begin{equation}
\label{eq:beamforming_cap}
 \begin{aligned}(\mathcal{P}1):\:\max_{\mathbf{\Theta,v}} &\log_2 \left( 1 + \frac{P}{\sigma^2}\mathbf{Hvv}^H\mathbf{H}\right)\\
 &||\mathbf{v}||^2\leq P,\:\:\mathbf{\Theta}\in\mathcal{T}
 \end{aligned}
\end{equation}
where $P$ is the total transmitted power. Due to the non-convexity of the problem we adopt an alternate optimization approach whereby we optimize $\mathbf{v}$ for a given $\mathbf{\Theta}$, and then optimize $\mathbf{\Theta}$ given $\mathbf{v}$. Assume that the transmitter knows $\mathbf{H}$ and computes its SVD, $\mathbf{H} = \mathbf{U} \bm{\Lambda} \mathbf{V}^H = \sum_i \lambda_i^{\frac12} \mathbf{u}_i \mathbf{v}_i^H$. If the receiver (transmitter) adopts $\mathbf{u}_j^H (\mathbf{v}_j$) as a unit-norm decoder (precoder), the maximum spectral efficiency is:
\begin{equation}
\label{eq:beamforming_cap}
    R_{\text{TxBF}} = \log_2 \left( 1 + \frac{P\lambda_{m}}{\sigma^2} \right),
\end{equation}
where $\lambda_m$ is the maximum eigenvalue of $\mathbf{H}^H \mathbf{H}$. To optimize \eqref{eq:beamforming_cap} with respect to $\mathbf{\Theta}$, a gradient-based iteration is proposed. Through the computation of the differential \cite{Hjrungnes2011ComplexValuedMD}:
\begin{equation}
d\lambda_{m} = \text{vec}^T \left( \frac{\mathbf{u}_m \mathbf{u}_m^H}{\mathbf{u}_m^H \mathbf{u}_m} \right) d\text{vec} \left( \mathbf{H}^H \mathbf{H} \right),
\end{equation}
and the differential $d\text{vec} \left( \mathbf{H}^H \mathbf{H} \right)$, we obtain:
\begin{flalign}
\label{eq:dVecJOK}
\begin{aligned}
        d\lambda_{m} &= \text{Re} \left[\text{vec}^T\left( \mathbf{H}_{ru}^T\mathbf{H}^*\mathbf{u}_m^*\mathbf{u}_m^T \mathbf{H}_{br}^T \right) \right ]  d \text{vec}(\mathbf{\Theta})\\
        &=\mathcal{D}_\mathbf{\Theta}\lambda_{m} \ d \text{vec}(\mathbf{\Theta})
\end{aligned}
\end{flalign}
from where the gradient row vector $\mathcal{D}_\mathbf{\Theta}\lambda_m$ can be read out.
\section{BD-RIS in Multimode MIMO Transmission}
\label{Optimizing BD-RIS in Multimode MIMO Transmission}
The received signal vector is now given by:
\begin{equation}
    \mathbf{y} = \mathbf{H} \mathbf{V}\mathbf{s} + \mathbf{n} \in \mathbb{C}^{K \times 1}
\end{equation}
where $\mathbf{s}$ contains the data stream, $\mathbf{V}$ is precoding matrix and $\mathbf{n}$ is the additive receiver Gaussian noise of zero mean and covariance $\sigma^2\mathbf{I}$. 
The optimization problem is formulated as:
\begin{equation}
\label{eq:MIMO_cap}
 \begin{aligned}(\mathcal{P}2):\:\max_{\mathbf{\Theta,R_s}} &\log_2 \det \left( \mathbf{I} + \frac{1}{\sigma^2} \mathbf{H}\mathbf{R_s} \mathbf{H}^H \right)\\
&\text{tr}(\mathbf{R_s})\leq P,\:\:\mathbf{R_s}\succeq0,\:\:\mathbf{\Theta}\in\mathcal{T}
 \end{aligned}
\end{equation}
where $\mathbf{R_s=V\Sigma V}^H$ and $\text{diag}(\mathbf{\Sigma})$ contains the powers of the elements of $\mathbf{s}$. To efficiently solve the non-convex problem we propose the alternate optimization between $\mathbf{R_s}$ and $\mathbf{\Theta}$.

Assume $\mathbf{H}$ is known, then $\mathbf{V}$ contains the right eigenvectors of $\mathbf{H}$ and the powers in $\mathbf{\Sigma}$ are obtained from the waterfilling solution. The scatter matrix of the RIS is optimized as:
\begin{equation}
\max_{\mathbf{\Theta}} \log_2 \det \left( \mathbf{I} + \frac{1}{\sigma^2} \mathbf{\bar{H}} \mathbf{\bar{H}}^H \right), \text{with} \ \mathbf{\bar{H}}=\mathbf{HV\Sigma}^{1/2}
\end{equation}

To obtain a gradient-based solution for $\mathbf{\Theta}$ let us compute the differential of the MIMO capacity as:
\begin{equation}
\label{eq:diferential_cap}
dC^u = d \left( \log_2 \det \left( \mathbf{I} + \frac{1}{\sigma^2} \mathbf{\bar{H}} \mathbf{\bar{H}}^H \right) \right)
\end{equation}
 with respect to $\mathbf{\Theta}$. Denote \( \mathbf{X} = \mathbf{I} + \frac{1}{\sigma^2} \mathbf{\bar{H}} \mathbf{\bar{H}}^H \), so \eqref{eq:diferential_cap} is \cite{Hjrungnes2011ComplexValuedMD}:
\begin{equation}
dC^u = \log_2 e \cdot \operatorname{Tr}(\mathbf{X}^{-1} d\mathbf{X}),
\end{equation}
where $d\mathbf{X} = \frac{1}{\sigma^2} \left(d\mathbf{\bar{H}} \mathbf{\bar{H}}^H + \mathbf{\bar{H}} d\mathbf{\bar{H}}^H \right)$. By using simple vectorization and trace properties, the differential boils down to:
\begin{equation}
\label{eq:dCOK}
\begin{aligned}
    dC^u &= 2\text{Re} \left[\text{vec}^T\left( \mathbf{H}_{ru}^T\mathbf{X}^{-1}\mathbf{\bar{H}}^*\mathbf{\bar{H}}_{br}^T \right)\right ] d \text{vec}(\mathbf{\Theta})\\&=\mathcal{D}_\mathbf{\Theta}C^u \ d \text{vec}(\mathbf{\Theta}).
    \end{aligned}
\end{equation}
from where the gradient vector $\mathcal{D}_\mathbf{\Theta}C^u$ can be read out and used in a gradient algorithm as described in the next section. With the RIS coefficients thus obtained, the decoder and precoder are taken as the matrices containing the left and right singular vectors of $\mathbf{H}$, and $\mathbf{P}$ is obtained from waterfilling.

\section{Constrained Gradient Ascend Iteration}
\label{Gradient ascend iteration}
We will use the gradients computed above to optimize the RIS scatter matrix iteratively. In passive D-RIS, $d\text{vec}(\mathbf{\Theta}) = j\text{diag}(\exp(j\phiM))d\phiM$, where the vector $\phiM$ contains all $N_r$ RIS phases.
For a passive BD-RIS, we need to ensure that \( \mathbf{\Theta}\in \mathcal{T} \) at each step of the iteration, so we face two approaches.
\subsection{Matrix exponentiation}
We can ensure symmetry and unitarity if $\mathbf{\Theta}=e^{j\mathbf{W}}$, with $\mathbf{W}\in\mathbb{R}^{N_r \times N_r}$ symmetric. Since $\mathbf{W}$ contains $N_r$ diagonal elements and $N_r(N_r-1)/2$ independent off-diagonal elements, we can express it in vectorized form as \cite{Hjrungnes2011ComplexValuedMD}:
\begin{equation}
\label{Wxy}
    \text{vec}(\mathbf{W}) = \mathbf{L}_d \mathbf{s} + (\mathbf{L}_u + \mathbf{L}_l) \mathbf{z},
\end{equation}
where $\mathbf{x} \in \mathbb{R}^{N_r}$ is a vector containing the diagonal elements of $\mathbf{W}$ and $\mathbf{z} \in \mathbb{R}^{N_r(N_r-1)/2}$ is a vector containing the unique off-diagonal elements. $\mathbf{L}_d \in \mathbb{R}^{N_r^2 \times N_r}$ is a selection matrix that extracts the diagonal elements in $\text{vec}(\text{diag}(\mathbf{W})) = \mathbf{L}_d \mathbf{s}$ and $\mathbf{L}_u, \mathbf{L}_l \in \mathbb{R}^{N^2 \times N(N-1)/2}$ are selection matrices that extract and place the upper and lower triangular parts of $\mathbf{W}$.

Since matrix exponentiation is defined as:
\begin{equation}
e^{j\mathbf{W}} = \sum_{k=0}^{\infty} \frac{(j\mathbf{W})^k}{k!}
\end{equation}
it can be checked that if $\mathbf{W} = \mathbf{U} \mathbf{\Lambda} \mathbf{U}^T$, where $\mathbf{U}$ is an orthogonal matrix and $\mathbf{\Lambda}$ is a diagonal matrix of real non-negative eigenvalues, then:
\begin{equation}
\label{eq:spectralDecomp}
\mathbf{\Theta} = e^{j \mathbf{W}} = \mathbf{U} e^{j \mathbf{\Lambda}} \mathbf{U}^T,
\end{equation}
and hence 
$\mathbf{\Theta}^H \mathbf{\Theta} = (\mathbf{U} e^{-j \mathbf{\Lambda}} \mathbf{U}^T)(\mathbf{U} e^{j \mathbf{\Lambda}} \mathbf{U}^T)= \mathbf{I}.$
From \eqref{eq:spectralDecomp}, $\mathbf{\Theta}$ is also symmetric. $d\text{vec}(\mathbf{\Theta})$ is a function of $d\mathbf{x}$ and $d\mathbf{y}$ and can be computed according to the lines described in \cite{Hjrungnes2011ComplexValuedMD} for patterned matrices as follows:
\begin{equation}
\label{eq:DWTheta1}
\mathcal{D}_\mathbf{W}\mathbf{\Theta} = j \sum_{k=0}^{\infty} \frac{1}{(k+1)!} \sum_{i=0}^{k} (j\mathbf{W}^T)^{k-i} \otimes (j\mathbf{W})^i.
\end{equation}
Using the eigenvalue decomposition of $\mathbf{W}$, $
\mathbf{W}^k = \mathbf{U} \mathbf{\Lambda}^k \mathbf{U}^T$, in \eqref{eq:DWTheta1} and operating the infinite summation we obtain:
\begin{equation}
\mathcal{D}_{\mathbf{W}}\mathbf{\Theta}=j (\mathbf{U} \otimes \mathbf{U}) \mathbf{S} (\mathbf{U} \otimes \mathbf{U})^T,
\end{equation}
where the diagonal matrix $\mathbf{S}$ of size $N_r^2\times N_r^2$ is built from $N_r$ diagonal matrices $\mathbf{S}=diag\left(\mathbf{S}_1,\dots,\mathbf{S}_{N_r}\right)$, each containing the following operation on the eigenvalues $\lambda$ of $\mathbf{W}$:
\begin{equation}
[\mathbf{S}_k]_{ll} = \begin{cases}
\frac{-j(e^{j\lambda_k} - e^{j\lambda_l})}{\lambda_k - \lambda_l}, & \lambda_k \neq \lambda_l \\
e^{j\lambda_k}, & \lambda_k = \lambda_l
\end{cases}
\end{equation}
where $k=1,\dots N_r$ and $l=1,\dots N_r$. Finally, we need to replace the differential of the scattering coefficients of the RIS:
\begin{equation}
    d\text{vec}(\mathbf{\Theta}) = j (\mathbf{U} \otimes \mathbf{U}) \mathbf{S} (\mathbf{U} \otimes \mathbf{U})^T d\text{vec}(\mathbf{W})
\end{equation}
in \eqref{eq:dVecJOK} and \eqref{eq:dCOK}. The term $d\text{vec}(\mathbf{W})$ as a function of $d\mathbf{x}$ and $d\mathbf{z}$ can be trivially obtained from \eqref{Wxy}.

\subsection{Projection on the unitary-symmetric manifold}
In this second approach we apply a two-step projection procedure. First, we compute the symmetric projection:
\begin{equation}
\mathbf{\Theta}_{\text{sym}} = \frac12(\mathbf{\Theta} + \mathbf{\Theta}^{T}).
\end{equation}
Then, we compute left and right singular vectors $\left [\mathbf{P},:,\mathbf{Q}\right]=\text{svd}(\mathbf{\Theta}_{sym})$ and construct the unitary and symmetric matrix as:
\begin{equation}
\mathbf{\Theta} = \mathbf{P} \mathbf{Q}^{H}.
\end{equation}
In practice, projection exhibits faster convergence than matrix exponentiation when inserted in a gradient iteration for $\mathbf{\Theta}$. Algorithm 1 shows the iterative gradient-based procedure where the projection approach is adopted, for the multi-mode MIMO transmission. The algorithm for beamforming is readily drawn.

\begin{algorithm}[h!]
\caption{Iterative optimization of BD-\ac{ris} coefficients}
\begin{algorithmic}
\State $t = 1$, initial random $\mathbf{\Theta}$
\State $J_0=0, J_1 = \epsilon$
\While {\text{$C^u$ is improved}}
\State $\mathbf{H} = \mathbf{H}_{bu}+\mathbf{H}_{ru} \mathbf{\Theta} \mathbf{H}_{br}$
\State$[\mathbf{U,\Lambda,V}]=\text{svd}(\mathbf{H}$)
\State$\mathbf{\Sigma}=\text{waterfilling}(\mathbf{\Lambda},P,\sigma^2)$
\State $\mathbf{\bar{H}}=\mathbf{HV\Sigma}^\frac12$, $\mathbf{\bar{H}}_{br}=\mathbf{H}_{br}\mathbf{V\Sigma}^\frac12$,$\mathbf{\bar{H}}_{bu}=\mathbf{H}_{bu}\mathbf{V\Sigma}^\frac12$
\While{abs$(J_t - J_{t-1})/J_t > \varepsilon$}
    \State $t \leftarrow t + 1$
    \State compute $D_\mathbf{\Theta} J$ \quad\quad\quad\% using eq. \eqref{eq:dCOK} 
    \State $\mathbf{\Theta}_t = \text{gradient\_update}(\mathbf{\Theta}_{t-1},D_{\mathbf{\Theta}} J) $
    \State $\mathbf{\Theta} = \frac12 (\mathbf{\Theta}_t+\mathbf{\Theta}_t^T)$
    \State $\left [\mathbf{P},:,\mathbf{Q}\right]=\text{svd}(\mathbf{\Theta})$
    \State $\mathbf{\Theta} = \mathbf{P}\mathbf{Q}^H$
    \State $\mathbf{\bar{H}}=\mathbf{\bar{H}}_{bu}+\Hru \boldsymbol{\Theta}\mathbf{\bar{H}}_{br}$
    \State compute $J_t=C^u$
\EndWhile
\EndWhile
\end{algorithmic}
\label{alg:alg1}
\end{algorithm}
In terms of the computational complexity of the inner loop as a function of $N_r$, transmit beamforming optimization requires a number of real products $\mathcal{O}(KN_r^2)$, our multi-mode MIMO capacity optimization needs $ \mathcal{O}(84 N_r^3)$ and the multi-mode MIMO approach in \cite{santamaria_c_bdris} requires $ \mathcal{O}(120 N_r^3)$. The later approach exhibits very close performance to ours.

\section{Results} 
\label{Results}
We evaluate here the effectiveness of gradient-based algorithms in optimizing the BD-RIS configurations. Specifically, we analyze how these algorithms enhance the spectral efficiency and coverage compared to conventional D-RIS architectures and we draw conclusions on their adaptability to different propagation environments.

To meet these goals, the UE is placed in every position of a $60\:m\times 60\:m$ area at a height of $1.5\:m$. The antenna gain at the UE (BS) is noted as $G_u (G_t)$. Moreover,  $\beta_{bu,k,l}$ in \eqref{eq:direct_channel_su} is calculated using the model in \cite{3gpp.38.901}[Table 7.4.1-1] for a micro scenario with $d_{k,m}\geq10$, considering that both BS and UE antennas are isotropic.
\begin{table}[b]
    \centering
    \begin{tabular}{|l|l|}
    \hline
        \textbf{Parameters} & \textbf{Values} \\ \hline
        Carrier Frequency, $f_c$ & 30 GHz  \\ \hline
        Antenna gains  & $G_t$  = $G_u$  = 3 dBi  \\ \hline
        Transmit power in DL, $P_t$ & 24 dBm  \\ \hline
        Receiver noise power & -94 dBm  \\ \hline
        Channel bandwidth, $B_w$ & 50 MHz\\  \hline
        BS location \& tilt (azim./elev.) & [30,60,10] m, $(\pi,\pi/2)$  \\ \hline
        UE height & 1.5 m \\ \hline
        RIS location \& tilt (azim./elev.) & [0,40,6] m, $(-\pi/2,\pi/2)$  \\ \hline
        Obstacle coordinates \& orientation & [23,40]-[33,40], along the $x$-axis      \\ \hline
    \end{tabular}
    \caption{System simulation parameters.}
    \label{system_param}
\end{table}
Following the lines in \cite{Ozdogan_IRSmodeling}, we use  \cite{balanis2012advanced}[Example 11-3] to find the channel gains of the compound channel for a reflecting RIS element of dimensions $a\times b$: 
\begin{gather}
    g_{ru,k,i}=\frac{G_u}{4\pi}\frac{ab}{d_{k,i}^2}\text{sinc}^2 (Y)\text{sinc}^2(W), \\
    g_{br,i,l}=\frac{G_t}{4\pi}\frac{ab}{d_{i,l}^2}\cos^2\psi_i, \\
    W = \frac{\pi a}{\lambda_c} \cos\theta_s,\:\:\:
    Y = \frac{\pi a}{\lambda_c} \left ( \sin\psi_i+\sin\theta_s\sin\psi_s\right),
\end{gather}
where $\psi_i$ corresponds to the azimuth angle of the incident wave on the RIS panel while $\psi_s$ and $\theta_s$ correspond to the azimuth and elevation angles of the scattered wave. The BS and the RIS are placed in coordinates [30,60,10] and [0,40,6] respectively, and are marked as red spots on top and left sides of the following figures. The elements of the RIS are distributed in $N_z=5$ rows and $N_y=N_r/5$ columns. We adopt $a=b=0.5\lambda_c$ for the RIS, and $M=4$ and $K=2$ antennas spaced by $0.5\lambda_c$ at the BS and UE respectively. Given these array dimensions, the Fraunhofer distance is $d_f=\frac{2D^2}{\lambda},$ where $D$ is the maximum dimension of the radiator. Having adopted $\lambda_c=0.01\:m$, $d_f$ for the RIS is around $800\: m$ so the studied area is well in near-field radiating conditions. Other system parameters are shown in Table \ref{system_param}. In all figures the spectral efficiency is in bits/s/Hz.

Given the non-convex nature of the BD-RIS coefficient optimization problems in the previous sections, advanced gradient-based iterations have been tested. Among these, the Momentum, Adam, and RMSprop optimizers have been tested for their robustnessa and convergence speed in navigating complex landscapes \cite{polyak1964some},\cite{kingma2014adam}. We observed that RMSprop performs better for the D-RIS optimization, while Adam shows superior performance when optimizing BD-RIS configurations. In practice, less than 70 iterations ensure stable convergence across all scenarios.

\subsection{BD-RIS Assistance vs. No BD-RIS Assistance}
\begin{figure}[h]
    \centering
    \begin{subfigure}{0.49\linewidth}
        \includegraphics[width=\linewidth]{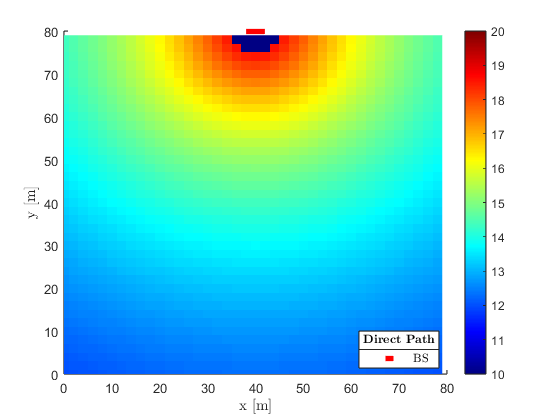}
        \caption{}
        \label{fig:MAP_DP}
        \end{subfigure}
    \begin{subfigure}{0.49\linewidth}
        \includegraphics[width=\linewidth]{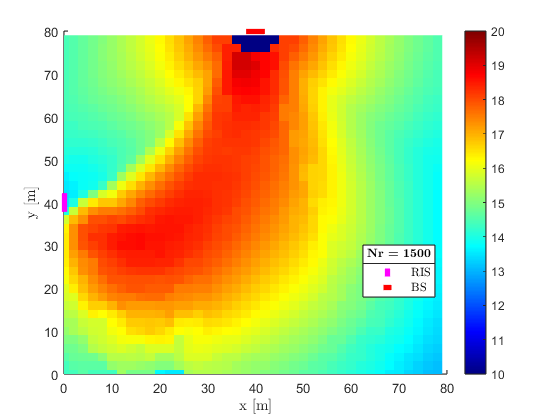}
        \caption{}
        \label{fig:MAP_BDRIS_1500}
    \end{subfigure}
    \caption{Spectral efficiency on the coverage area for multi-mode MIMO transmission: (a) no RIS assistance, (b) with optimized BD-RIS of $N_r=1500$.}
\end{figure}

Results comparing spectral efficiency in different scenarios (direct path in Fig. \ref{fig:MAP_DP}, BD-RIS optimized in Fig. \ref{fig:MAP_BDRIS_1500}) reveal insights regarding performance improvement. The multi-mode MIMO capacity $C^u$ is noticeably increased in a large area, specially near the BD-RIS optimized for MIMO capacity.

\begin{figure}
    \centering
    \begin{subfigure}{0.49\linewidth}
        \includegraphics[width=\linewidth]{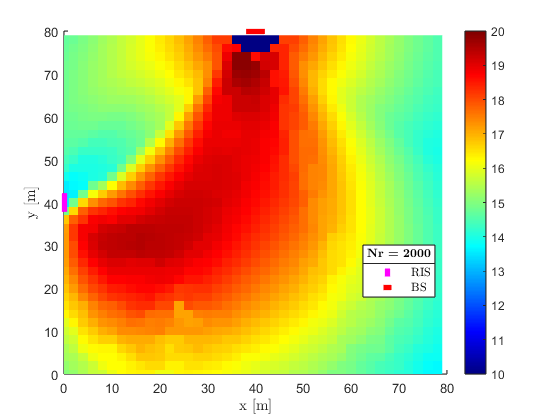}
        \caption{}
        \label{fig:MAP_BDRIS_2000}
    \end{subfigure}
    \begin{subfigure}{0.49\linewidth}
        \includegraphics[width=\linewidth]{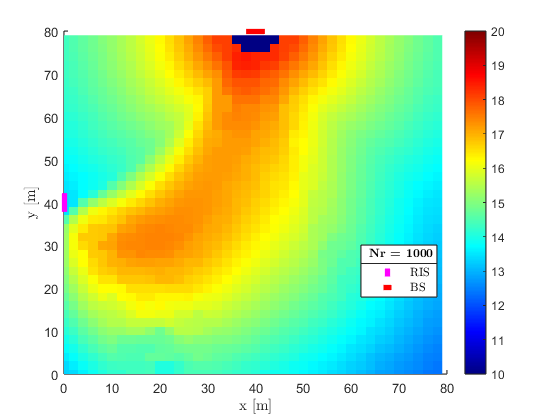}
        \caption{}
        \label{fig:MAP_BDRIS_1000}
    \end{subfigure}
    \caption{Spectral efficiency for multi-mode MIMO with optimized BD-RIS assisted transmission with (a) $N_r=2000$, (b) $N_r=1000$.}
\end{figure}

\begin{figure}[t]
\captionsetup{justification=centering,margin=0.25cm}
    \begin{center}
        \includegraphics[scale=0.34]{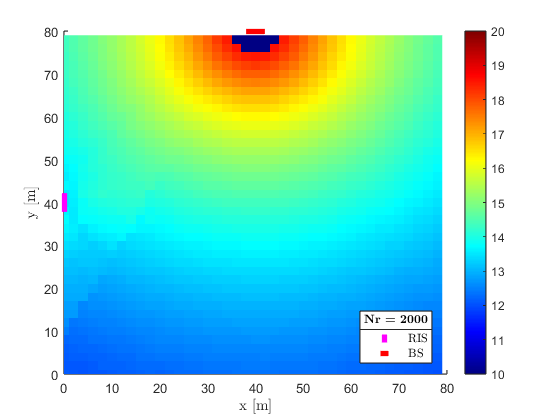}
        \caption{Spectral efficiency of the TxBF on a coverage area for a DL MIMO transmission assisted by a BD-RIS with $N_r=2000$.}
        \label{fig:MAP_BDRIS_2000_TxBF}
    \end{center}
\end{figure}

Figs. \ref{fig:MAP_BDRIS_2000}, \ref{fig:MAP_BDRIS_1000} show that the peak spectral efficiency for the MIMO capacity is consistently higher as $N_r$ increases for optimized RIS.
Interestingly, we observe minimal improvement in the TxBF optimized BD-RIS case for $N_r=2000$ (compare Fig. \ref{fig:MAP_BDRIS_2000_TxBF} and Fig. \ref{fig:MAP_DP}): while multi-mode MIMO capacity focuses on maximizing the overall data rate and the RIS is able to enhance channel matrix rank, TxBF optimization aims to align the transmitted signal with the dominant channel mode which is challenging for high path loss as the RIS has limited ability to contribute to significant SNR improvement. This can be checked in Table \ref{tab:cdf_RIS_vs_BDRIS_C_all} for different number of RIS elements: when $N_r$ is increased, TxBF gains are not significant.

\subsection{D-RIS vs. BD-RIS}
\begin{figure}[h]
    \captionsetup{justification=centering,margin=0.25cm}
    \begin{center}
        \includegraphics[scale=0.34]{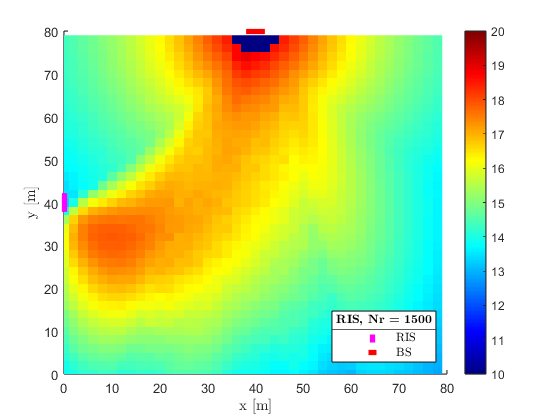}
        \caption{Spectral efficiency on a coverage area for a MIMO capacity transmission assisted by a D-RIS with $N_r=1500$.}
        \label{fig:MAP_RIS_1500}
    \end{center}
\end{figure}

We focus on analyzing the impact of deploying BD-RIS versus conventional D-RIS in spectral efficiency. The comparison of Figs \ref{fig:MAP_BDRIS_1500} and \ref{fig:MAP_RIS_1500} reveals that BD-RIS outperforms D-RIS as a result of the increased degrees of freedom offered by the BD-RIS (see also Table \ref{tab:cdf_RIS_vs_BDRIS_C_all}). We observe that the gain scales with the number of RIS elements as the gap in the maximum capacity achieved between the BD-RIS and the D-RIS increases, thus emphasizing the potential of BD-RIS.

\begin{table}[h]
\centering
\begin{tabular}{c|cc|cc|}
\cline{2-5}
                                                          & \multicolumn{2}{c|}{\textbf{Peak spectral eff}}      & \multicolumn{2}{c|}{\textbf{Gain wrt DP}} \\ \cline{2-5} 
                                                          & \multicolumn{1}{c|}{\text{MIMO cap}} & \text{TxBF} & \multicolumn{1}{c|}{\text{MIMO cap}}  & \text{TxBF}  \\ \hline
\multicolumn{1}{|c|}{\text{Direct Tx (DT)}}                & \multicolumn{1}{c|}{14.22}     & 14.2203       & \multicolumn{1}{c|}{-}      & -        \\ \hline
\multicolumn{1}{|c|}{\text{$N_r$ = 1000 D-RIS}}      & \multicolumn{1}{c|}{16.98}     & 14.2210       & \multicolumn{1}{c|}{19.40\%}     & 0.011\%        \\ \hline
\multicolumn{1}{|c|}{\text{$N_r$ = 1000 BD-RIS}}  & \multicolumn{1}{c|}{17.47}     & 14.2216       & \multicolumn{1}{c|}{22.88\%}     & 0.009\%        \\ \hline
\multicolumn{1}{|c|}{\text{$N_r$ = 1500 D-RIS}}      & \multicolumn{1}{c|}{17.87}     & 14.2228       & \multicolumn{1}{c|}{25.63\%}     & 0.017\%        \\ \hline
\multicolumn{1}{|c|}{\text{$N_r$ = 1500 BD-RIS}}  & \multicolumn{1}{c|}{18.65}     & 14.2231       & \multicolumn{1}{c|}{31.16\%}     & 0.015\%        \\ \hline
\multicolumn{1}{|c|}{\text{$N_r$ = 2000, D-RIS}}     & \multicolumn{1}{c|}{18.42}     & 14.2237       & \multicolumn{1}{c|}{29.54\%}     & 0.024\%        \\ \hline
\multicolumn{1}{|c|}{\text{$N_r$ = 2000, BD-RIS}} & \multicolumn{1}{c|}{19.49}     & 14.2239       & \multicolumn{1}{c|}{37.06\%}     & 0.023\%        \\ \hline
\end{tabular}
\caption{Evaluation of achieved peak spectral efficiency in the area, for direct BS-UE transmission (DT), D-RIS assisted transmission and BD-RIS assisted transmission, optimized for multi-mode MIMO capacity and TxBF.}
\label{tab:cdf_RIS_vs_BDRIS_C_all}
\end{table}
\subsection{Coverage behind obstacles}
In this section we evaluate the achieved rates in both MIMO configurations (TxBF and multi-mode MIMO transmission) in a scenario with an obstacle obstructing the direct path with a 10 dB attenuation, resulting in non-line-of-sight (NLoS) conditions. To do so, we incorporate a 10 meters wide obstacle (marked as a black line in the following plots).
While TxBF is generally effective in enhancing the signal power towards intended directions, its efficacy diminishes in NLoS conditions. Conversely, multi-mode MIMO offers improved flexibility in distributing power across modes thus enhancing performance.

\begin{figure}[h]
    \centering
    \begin{subfigure}{0.49\linewidth}    
        \includegraphics[width=\linewidth]{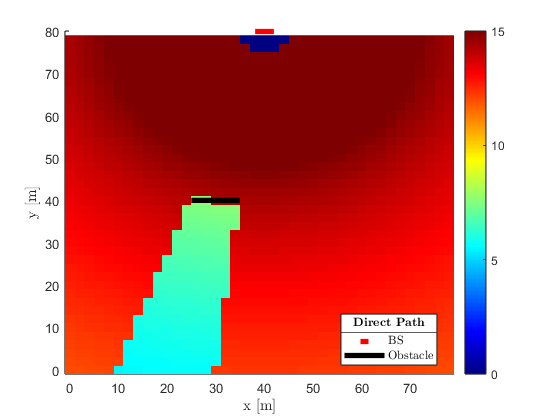}
        \caption{}
        \label{fig:MAPA_DP_OBS}
        \end{subfigure}
    \begin{subfigure}{0.49\linewidth}    
        \includegraphics[width=\linewidth]{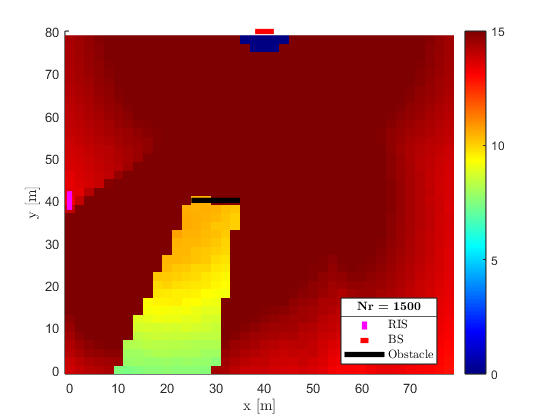}
        \caption{}
        \label{fig:MAPA_RIS_OBS}
    \end{subfigure}
    \vfill
    \begin{subfigure}{0.49\linewidth}
    
        \includegraphics[width=\linewidth]{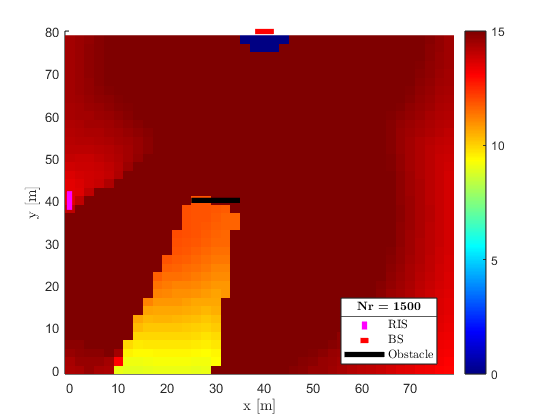}
        \caption{}
        \label{fig:MAPA_BDRIS_OBS}
    \end{subfigure}
    \caption{Spectral efficiency when an obstacle is present: (a) no RIS assisted downlink transmission, (b) BD-RIS assisted with TxBF, (c) BD-RIS assisted with multi-mode MIMO transmission. $N_r=1500$ in the later cases.}
\end{figure}

\cref{fig:MAPA_BDRIS_OBS,fig:MAPA_RIS_OBS,fig:MAPA_DP_OBS} reveal that the BD-RIS configuration significantly enhances peak spectral efficiency compared to direct path transmission in the shadowed region behind the obstacle. For instance, when $N_r = 1500$, the peak spectral efficiency in the multibeam transmission case increases by approximately $62\%$ relative to the direct path scenario. Conversely, the transmit beamforming case achieves a modest increase of $12\%$, yet it still demonstrates some improved performance compared to the scenario without obstacles. This result highlights the effectiveness of BD-RIS for coverage improvement.

\begin{table}[h]
\centering
\begin{tabular}{c|cc|cc|}
\cline{2-5}
                                                         & \multicolumn{2}{c|}{\textbf{Peak spectral eff}}      & \multicolumn{2}{c|}{\textbf{Gain wrt DP}} \\ \cline{2-5} 
                                                         & \multicolumn{1}{c|}{\text{MIMO cap}} & \text{TxBF} & \multicolumn{1}{c|}{\text{MIMO cap}}  & \text{TxBF}  \\ \hline
\multicolumn{1}{|c|}{\text{Direct Tx (DT)}}               & \multicolumn{1}{c|}{5.51}      & 5.50        & \multicolumn{1}{c|}{-}      & -        \\ \hline
\multicolumn{1}{|c|}{\text{$N_r$ = 1000 D-RIS}}     & \multicolumn{1}{c|}{6.48}      & 5.82       & \multicolumn{1}{c|}{17.69\%}     & 5.73\%        \\ \hline
\multicolumn{1}{|c|}{\text{$N_r$ = 1000 BD-RIS}} & \multicolumn{1}{c|}{7.80}     & 5.79       & \multicolumn{1}{c|}{41.63\%}     & 5.28\%        \\ \hline
\multicolumn{1}{|c|}{\text{$N_r$ = 1500 D-RIS}}     & \multicolumn{1}{c|}{7.43}      & 6.21        & \multicolumn{1}{c|}{35.03\%}     & 12.80\%       \\ \hline
\multicolumn{1}{|c|}{\text{$N_r$ = 1500 BD-RIS}} & \multicolumn{1}{c|}{8.91}     & 6.17       & \multicolumn{1}{c|}{61.90\%}     & 12.07\%       \\ \hline
\end{tabular}
\caption{Evaluation the achieved peak spectral efficiency in the area behind the obstacle for direct BS-UE transmission (DT), D-RIS vs. BD-RIS.}
\label{tab:OBS_RIS_vs_BDRIS_C}
\end{table}

\section{Conclusions}
\label{Conclusions}
The results confirm that optimizing the BD-RIS configurations leads to a significant capacity enhancement compared to conventional D-RIS, taking advantage of the additional degrees of freedom, under both LOS and NLOS conditions. A key finding is that multi-mode transmission (from MIMO capacity optimization) benefits further from RIS assistance than transmit beamforming. This trend is particularly evident in challenging environments with shadowed areas.
In general, results highlight the critical role of BD-RIS in next-generation wireless networks, demonstrating its potential to improve spectral efficiency and coverage. Future work focuses on reducing complexity and address scalability in other MIMO scenarios.

\bibliographystyle{IEEEtran} 
\bibliography{mybiblio}

\end{document}